\documentclass[12pt]{iopart}
\usepackage{iopams}  
\def\Div{\hbox{\rm Div}}
\def\Lie{\pounds}
\def\del{\partial}                   
\def\arr{\rightarrow }            

\def\E{\Bbb{E}}
\def\F{\Bbb{F}}
\def\R{\Bbb{R}}
\def\C{\Bbb{C}}

\def\calP{\mathcal{P}} 
\def\calC{\mathcal{C}} 
\def\calE{\mathcal{E}} 
\def\calW{\mathcal{W}} 
\def\calL{\mathcal{L}} 
\def\calA{\mathcal{A}} 

\def\La{\Lambda}
\def\na{\nabla}
\def\la{\lambda}
\def\Si{\Sigma}
\def\si{\sigma}

\def\ep{\epsilon}
\def\al{\alpha}
\def\be{\beta}
\def\Ga{\Gamma}
\def\ga{\gamma}

\def\de{\delta}

\def\vp{\varphi}

\def\Aut{\hbox{\rm Aut}}
\def\dim{\hbox{\rm dim}}
\def\ds{\hbox{\bf ds}}
\def\Spin{\hbox{\rm Spin}}
\def\GL{\hbox{\rm GL}}
\def\SO{\hbox{\rm SO}}
\def\d{\hbox{\rm d}}

\def\BE{\begin{equation}}
\def\EE{\end{equation}}
\def\Ref#1{(\ref{#1})}

\begin{document}

\title[SUSY and gauge natural theories]{Supersymmetries and Gauge Natural Theories}

\author{L.Fatibene, M.Francaviglia}

\address{Dipartimento di Matematica, Universit\`a degli Studi di Torino, 
Via Carlo Alberto 10, 10123 Torino, Italy}

\begin{abstract}
We discuss the gauge natural formulation of supersymmetric theories and supergravity,
with the aim to show that the standard and the supersymmetric frameworks admit in fact a
unifying mathematical language.

\end{abstract}



\maketitle

\section{Introduction}
Gauge natural theories have recently been shown to encompass a
wide class of physically relevant field theories (see e.g.\ ref. \cite{Kolar}--\cite{Ray} for definitions and applications). 
In fact, a fundamental field theory is necessarily {\it gauge natural}, especially when gauge transformations and/or spinors
are involved (see \cite{Spinors}).

The gauge natural formalism consists basically in assigning a suitable principal bundle  
$\calP=(P,M,p;G)$, called the {\it structure bundle} of the theory, and assuming that
symmetries can be identified with
the automorphisms group $\Aut(\calP)$.
Then the bundle  of physical configurations is necessarily a gauge natural bundle
$\calC=(C,M,\pi;F)$ associated to $\calP$  via some suited group representation
(see \cite{Kolar}, \cite{Bologna}). 
Accordingly, $\Aut(\calP)$ acts canonically on $\calC$,
{\it (pure) gauge transformations} are {\it vertical}, while
{\it generalized gauge transformations} are general automorphisms acting on
$\calC$ by means of the canonical action.

Gravity (in its purely metric, metric-affine and tetrad formulations with or without a reference
background), its interaction with  Yang-Mills fields as well as Bosonic and Fermionic matter, is most
suitably described in the gauge natural framework. 
Moreover, gauge natural theories allow a complete, general and satisfactory framework for
conserved quantities. 
All gauge natural theories admit in fact {\it superpotentials}, i.e.\ their N\"other currents are exact
regardless of the topology of spacetime $M$ and there exist a basically canonical choice of the
appropriate potential, which is usually called  a {\it superpotential} (see \cite{Remarks}, \cite{Robutti}
for details).

{\it Supersymmetric theories} and {\it supergravity
models} can in fact be included in the gauge natural framework, even if
supersymmetries have a number of features which set them definitely out of the standard framework for
ordinary symmetries.
First of all, sypersymmetries usually leave the Lagrangian invariant modulo pure divergence terms and
sometimes only on-shell (i.e.\ modulo terms which vanish along solutions).
Secondly, supersymmetries are not transformations on the configuration bundle, since
$1$-parameter deformations of fields do not depend just on fields, but also
on derivatives up to some (finite) order $k>0$ (usually $k=1$). 
Thirdly, in supersymmetric models spinor fields are anticommuting in view of a path-integral
quantization. 
Finally, the parameter of the supersymmetries is an anticommuting spinor so that
supersymmetries span a superalgebra rather than an ordinary Lie algebra. 
Here we want to give the necessary guidelines needed to enlarge the standard framework to encompass
these features under a unifying geometrical and global viewpoint. 
Further details may be found in \cite{Ray}, \cite{Libro}, \cite{SuperTesi}.

\section{Generalized N\"other theorem}

Let $\calC=(C,M,\pi;F)$ be the configuration bundle of a physical field theory and $(x^\mu, y^i)$ be
local fibered coordinates on it. Here $x^\mu$ are coordinates on spacetime $M$ and
$y^i$ are coordinates on the standard fiber $F$ (i.e.\ the fields).
A Lagrangian $L$ is given in terms of fields and their partial derivatives up to some finite order, say
$k$. 
The Lagrangian is mathematically described as an object living on the so-called {\it jet-bundle}
$J^k\calC$ having natural fibered coordinates $(x^\mu, y^i, y^i_\mu,\dots,  y^i_{\mu_1\dots\mu_k})$,
where $y^i_{\mu_1\dots\mu_s}$ ($1\le s\le m=\dim(M)$)
are symmetric w.r.t. $(\mu_1\dots\mu_s)$ since they represent the derivatives.

More precisely, $L$ is a horizontal form over $J^k\calC$, i.e.\ it is of the
following form 
\BE
L=\calL(x^\mu;y^i, y^i_\mu,\dots,  y^i_{\mu_1\dots\mu_k})\ds\>\>,
\qquad
\ds=\d x^1\land\d x^2\land\dots\land\d x^m
\EE
An {\it infinitesimal transformation} is a vector field $\Xi$ over $\calC$ which projects
over a vector field $\xi$ over $M$ and lifts in an obvious way to a vector field $j^k\Xi$ over $J^k\calC$.
The {\it Lie derivative of a section $\si:M\arr C$} along $\Xi$ is defined by
\BE
\Lie_\Xi\si=T\si(\xi) - \Xi \circ \si=(\Lie_{\Xi}y^i)\>\del_i\equiv (\de y^i)\>\del_i
\label{LieDer}\EE
We say that the Lagrangian is {\it covariant} w.r.t. the infinitesimal transformation $\Xi$
(or, equivalently, that $\Xi$ is an {\it infinitesimal symmetry} for $L$) when
the so-called {\it covariance identity} holds:
\BE
\Div(i_\xi\> L)= <\de L\>|\> j^k\Lie_\Xi \si>
\label{CovId}\EE
where $\Div$ is the {\it formal divergence operator} and $<\de L\>|\> j^k\Lie_\Xi \si>$ is given by 
\BE
<\de L\>|\> j^k\Lie_\Xi \si>=
{\del\calL\over \del y^i}\de y^i+
{\del\calL\over \del y^i_\mu}\del_\mu\de y^i+\dots
{\del\calL\over \del y^i_{\mu_1\dots\mu_k}}\del_{\mu_1\dots\mu_k}\de y^i
\EE

By a canonical covariant integration by parts on the r.h.s. of \Ref{CovId}, we obtain globally:
\BE
<\de L\>|\> j^k\Lie_\Xi \si>\>=\><\E_L\>|\> \Lie_\Xi \si>+\>\Div\><\F_L\>|\> j^{k-1}\Lie_\Xi \si>
\EE
where $\E_L$ are field equations and $\F_L$ collects all the ensuing pure divergence terms. 
Both $\E_L$ and $\F_L$ are global horizontal forms over the appropriate prolongations of
$\calC$.

It is known that the standard {\it N\"other theorem} $\Div\>\calE=\calW$ holds,
where one sets
\BE
\cases{
\calE_L=<\F_L\>|\> j^{k-1}\Lie_\Xi \si>-i_\xi\> L\cr
\calW_L=-<\E_L\>|\> \Lie_\Xi \si>\cr
}
\EE
Since $\calW_L=0$ along solutions, the {\it N\"other current $\calE_L$} is conserved 
on-shell.

If the Lagrangian is covariant modulo a pure divergence term, i.e.
the {\it generalized covariant identity} holds true
\BE
\Div(i_\xi\> L)= <\de L\>|\> j^k\Lie_\Xi \si> +\> \Div\>\al
\label{GenCovId}\EE
for some horizontal form $\al$, then a {\it generalized N\"other theorem} holds with
a new prescription for the {\it generalized N\"other current}, namely
\BE
\hat\calE_L=<\F_L\>|\> j^{k-1}\Lie_\Xi \si>-i_\xi\> L +\al
\label{GenNoeCurr}\EE
i.e.\ $\Div\>\hat\calE_L=0$ still holds on-shell.

\section{Generalized vector fields}

The standard N\"other theorem relies on a number of
unessential hypotheses that can be relaxed.
As we discussed above the first is {\it exact covariance}.
Another is the following:
a vector field $\Xi$ over $\calC$ is actually  used to Lie-drag sections.
However, not all one-parameter families of sections are generated in this way
and and in fact N\"other theorem just needs a one-parameter family such that
\Ref{GenCovId} holds true.

Lie-dragging can be defined along objects more general  than vector fields,
i.e. along so-called {\it generalized vector fields} (see \cite{Ray}, \cite{Olver}).
We shall here restrict to generalized vertical vector fields of the form
\BE
X=X^j(x^\mu, y^i, y^i_\mu,\dots,y^i_{\mu_1\dots\mu_k})\del_j
\EE
which of course are not vector fields on $\calC$ since they depend on the derivatives of fields.
Generalized vector fields are in fact sections $X$ of a
bundle $\pi^\ast:(\pi^k_0)^\ast V\calC\arr J^k\calC$ suitably defined as a {\it pull-back} 
of the so-called vertical bundle $V\calC$ (see \cite{Ray}, \cite{Olver}).

For each section $\si$ the composition $X_\si= X\circ j^k\si:M\arr (\pi^k_0)^\ast V\calC$ is then a
{\it vertical vector field} defined on the image of $\si$ and thence it can be used to drag the
section itself. 
The Lie-dragging is still defined by equation \Ref{LieDer},
covariance by \Ref{GenCovId} and N\"other theorem follows again by prescription
\Ref{GenNoeCurr}.\footnote{ 
In principle one could also decide to drag sections directly on $J^k\calC$
rather than  restricting {\it a priori} to prolongations of Lie-draggings on $\calC$.
However, in most cases this procedure is not more general, since the transformations turn out {\it a
posteriori} to be prolongations in order to preserve the the fundamental structure of $J^k\calC$ (see
e.g.\ \cite{Olver}).

Another equivalent viewpoint is to consider all trasformations as given on the infinite jet
bundle $J^\infty\calC$, i.e.\ the {\it inverse limit} of the family of all $J^k\calC$
(see \cite{Olver}). 
}

\section{Anticommuting spinors}

Let us denote by $\Spin(r,s)$ the spin group of signature $(r,s)$ ($r+s=m=\dim(M)$)
and by $\GL(m)$ the general linear group.
We denote by $\ell:\Spin(r,s)\arr \SO(r,s)$ the covering map exhibiting the spin group as a double
covering of the corresponding orthogonal group.
Moreover, let us denote by $L(M)$ the principal bundle of linear frames on $M$, which has
$\GL(m)$ as a structure group.

Spinors can be defined on a general spin-manifold $M$  (see \cite{Lawson}) which shall be here assumed
to admit global metrics of the given signature.
 The first step is to regard spin structures on $M$ as  dynamical variables by introducing as follows
the so-called {\it spin frames} (see \cite{Spinors}, \cite{Polonia}, \cite{Bologna}). 
We fix a principal bundle $\Si$ with $\Spin(r,s)$  as structure group; spin frames
are principal morphisms $e:\Si\arr L(M)$. We consider the action on $\GL(m)$
\BE
\la:\GL(m)\times \Spin(r,s) \times \GL(m)\arr \GL(m):
(J, S, e)\mapsto \ell(S)\cdot e\cdot J^{-1}
\EE
and we define the associated bundle $\Si_\la=(L(M)\times_M \Si)\times_\la \GL(m)$;
it has fibered coordinates $(x^\mu, e^a_\mu)$ and it is thence locally analogue to the frame bundle.
For this reason spin frames are often identified in literature with standard frames ({\it vielbein}).
We stress however that these similarities have only a local nature since spin frames and ordinary frames
have completely different behaviours w.r.t. Lie-dragging and covariant derivative (see \cite{Spinors}).
The bundle $\Si_\la$ is by construction a gauge natural bundle and its sections
are in one-to-one correspondence with spin frames.

Spin frames can be defined on any spin manifold (even on non-parallelizable ones)
and each spin frame induces a metric by $g_{\mu\nu}= e^a_\mu\>\eta_{ab}\> e^b_\nu$
where $\eta_{ab}$ is the standard diagonal matrix with signature $(r,s)$.

Supersymmetries strongly rely on anticommuting spinors. 
Naively speaking, if $\psi^i$ denote the components of a spinor $\psi$ then quadratic forms such as
$\psi^i\vp^j$ are assumed to be antisymmetric in the exchange of the spinors, i.e.
$\psi^i\vp^j=-\vp^j\psi^i$, in view of a path integral quantization of the field theory
(according to Fermi-Dirac quantization rules for Fermions).
Mathematically, this behaviour is implemented by assuming that each component $\psi^i$ takes its
value in the odd part of the (complexified) exterior algebra $\La_-(W)\otimes\C$ of some suitable
vector space $W$. Let us denote by $V=[\La_-(W)\otimes \C]^n$ the space so generated;
since $\La_-(W)\otimes \C$ is a complex vector space of dimension $2^{\rm{dim}(W)-1}$, $V$ is a
complex vector space and we are entitled to use anticommuting coordinates $\psi^i$ in $V$.
Each $\psi^i$ is actually a block of $2^{\rm{dim}(W)-1}$ (complex) ordinary coordinates.

The spin group $\Spin(r,s)$  acts as follows on spinors by block matrix multiplication on
anticommuting coordinates:
\BE
\rho:\Spin(r,s)\times V\arr V:(S, \psi)\mapsto S^i_{\>j}\>\psi^j
\EE
Thence we can define the associated bundle $\Si_\rho: \Si\times_\rho V$
the sections of which are called {\it anticommuting spinors}.

Choosing the matrix representation of the spin group is equivalent to fix a set of Dirac matrices $\ga_a$
($a=1,\dots,m$) such that $\ga_a\ga_b+\ga_b\ga_a=2\eta_{ab}$.
The charge conjugation operator is defined as the element $C$ in the appropriate Clifford algebra such
that $C\ga_a C^{-1}= -^t\ga_a$, where $^t\ga_a$ is the transpose of $\ga_a$.
The {\it charge conjugated spinor} is defined as $\phi^c=C({}^t\bar\psi)$ where
$\bar\psi=\psi^\dagger\ga_0$ is the {\it adjoint spinor}, and $\psi^\dagger$ is the transpose
conjugate spinor. Spinors which are self charge conjugated ($\psi^c=\psi$) are called {\it Majorana
spinors}, provided they exist (e.g., this happens in $4$-dimension Lorentzian spacetimes). 
In this case the spinor space $V$ splits as
$V=V_+\oplus V_-$.  The action of the spin group on $V$ preserves this splitting and thence also the bundle
$\Si_\rho$ splits as $\Si_\rho=\Si^+_\rho\oplus \Si^-_\rho$.
Sections of $\Si^+_\rho$ are in one-to-one correspondence with Majorana spinors.

\section{Wess-Zumino model}

Wess-Zumino model is the simplest supersymmetric model (see \cite{Wess}).
It will be used here as an example to show that the gauge natural framework is suited to deal with all
supersymmetric theories.  
We shall consider it on a general spin manifold $M$ with a fixed spin bundle
$\Si$. Fields are: a spin frame $e^a_\mu$, an anticommuting Majorana spinor $\psi$ and four scalar
densities
$(A,B,C,D)$ of arbitrary weights $(\al,\be,\ga,\de)$.
Each scalar density, say $A$, is a section of a natural bundle $\calA^\al=L(M)\times_{\la_\al} \R$
associated to $L(M)$ by means of the representation
\BE
\la_\al:\GL(m)\times \R\arr \R: (J, A)\mapsto (\det J)^{-\al}\>A
\EE  
Fibered coordinates on $\calA^\al$ are thence $(x^\mu, A)$.

The configuration bundle of the Wess-Zumino model is thence
\BE
\calC_{WZ}= \Si_\la\times_M \Si^+_\rho\times_M \calA^\al\times_M \calA^\be\times_M \calA^\ga\times_M \calA^\de
\EE
which is a gauge natural bundle associated to the structure bundle $\Si$
and it has fibered ``coordinates'' $(x^\mu, e^a_\mu, \psi^i, A, B, C, D)$.
We recall that strictly speaking these are not coordinates since the $\psi^i$ anticommute.

We consider the Lagrangian
\BE
\eqalign{
L_{WZ}=&\hbox{$1\over 2$}\left[(\na_a A)( \na^a A) \>e^{2\al}+ D^2\>e^{2\de}  
+ 2mAD \> e^{\al+\de}\right]e\>\ds +\cr
&+\hbox{$1\over 2$}\left[(\na_a B)( \na^a B) \>e^{2\be} + C^2\>e^{2\ga}   -2 m BC\>e^{\be+\ga}
\right]e\>\ds\cr
&-\bar\psi\left(i\ga^a\na_a\psi+m\psi\right)e\>\ds\cr 
}
\label{WZL}\EE
where $e$ denotes the determinant of the spin frame $e^a_\mu$, $e_a^\mu$ is the inverse of $e^a_\mu$,
the covariant derivatives of the scalar densities are defined as usual, e.g. $\na_a A=e_a^\mu(\d_\mu A
-\al\Ga^\la_{\>\la\mu} A)$;
$\Ga^\al_{\>\be\mu}$ (resp. $\Ga^{ab}_\mu$) is the Levi-Civita connection of
the metric $g$ (resp. the spin connection) uniquely induced by the spin frame $e^a_\mu$ and the covariant
derivative of the spinor is
\BE
\na_a\psi= e_a^\mu(\d_\mu\psi +\hbox{$1\over 8$}\Ga^{ab}_{\>\mu}[\ga_a, \ga_b]\psi)
\EE

Since $\calC_{WZ}$ is gauge natural all automorphisms of $\Si$ act canonically on $\calC_{WZ}$.
The Lagrangian \Ref{WZL} is covariant with respect to generalized gauge transformations
and consequently the theory is gauge natural in the sense of \cite{Remarks}.
One can thence canonically define conservation laws which admit superpotentials, as proved in
general for any gauge natural theory (see \cite{Bologna}).

\section{Supersymmetry transformations}

Supersymmetries can be now introduced as generalized vector fields.
Let $\ep$ be a  covariantly constant ($\na_\mu\ep=0$) Majorana spinor.
Let us consider the generalized vector field 
\BE
X=\de e^a_\mu {\del\over \del e^a_\mu} 
+\de \psi {\del\over \del \phi} 
+ \de A{\del\over \del A}
+ \de B{\del\over \del B}
+ \de C{\del\over \del C}
+ \de D{\del\over \del D}
\label{GenSuSy}\EE
where we set $\de e^a_\mu=0$ and
\BE
\cases{
\matrix{
\de A= \hbox{$a\over 2$}(\bar\ep \psi) \>e^{-\al} \hfill          
&,&& \de C= -\hbox{$a\over 2$}(\bar\ep \ga^5\ga^a\na_a\psi)\>e^{-\ga}\hfill\cr
\de B= -i\hbox{$a\over 2$}(\bar\ep \ga^5\psi)\> e^{-\be}\hfill 
&,&& \de D= i\hbox{$a\over 2$}(\bar\ep \ga^a\na_a\psi)\> e^{-\de}\hfill\cr
}\cr
\de\psi=\hbox{$a\over 2$}\left[i(\ga^a\ep)\> \na_a A\>e^{\al} +(\ga^a\ga^5\ep)\> \na_aB \>e^{\be} +
i(\ga^5\ep)\>C\>e^{\ga} + (\ep)\> D\>e^{\de}\right]\cr
}
\EE
Notice that \Ref{GenSuSy} is a generalized vector field since it depends on the derivatives of fields.

One can easily prove that these infinitesimal transformations are symmetries for the Lagrangian \Ref{WZL},
modulo the divergence term
\BE
\eqalign{
\al=&\hbox{$a\over 4$}\Big[\Big(
2im A (\bar\ep\ga^\mu\psi)
+2\na^\mu A (\bar\ep\psi)
-\na_\nu A (\bar\ep\ga^\nu\ga^\mu\psi)\Big)\>e^{\al+1}
+\cr
&+\Big(2m B (\bar\ep\ga^5\ga^\mu\psi)
-2i\na^\mu B (\bar\ep\ga^5 \psi)
+i\na_\nu B (\bar\ep\ga^5\ga^\nu\ga^\mu\psi)\Big)\>e^{\be+1}+\cr
&+iD (\bar\ep\ga^\mu\psi) \>e^{\de+1}
+C (\bar\ep\ga^\mu\ga^5\psi)\>e^{\ga+1}
\>\Big]\ds_\mu\cr
}
\EE
where $\ds_\mu=i_{\del_\mu}\ds$ is the $(m-1)$-surface local volume.

\section{Conclusions and Perspectives}

First of all we have shown on a simple example that gauge natural theories may admit supersymmetries which are
generalized symmetries.
Our example can be easily coupled to the Hilbert Lagrangian 
$L= R^{ab}_{\>\>\mu\nu} e_a^\mu e_b^\mu e\>\ds$ for the spin frame which describes interactions with
gravity. The theory is {\it necessarily} a gauge natural theory since spin frames are truly gauge natural
(and they have to be used since ordinary frames or vielbeins globally fail in replacing
spin frames in a generic topology for $M$).

Another important example of supersymmetric theory is the
Rarita-Schwinger (RS) model which deals with spin frames $e^a_\mu$ and a $3/2$-spinor field $\psi^i_\mu$,
i.e.\ sections of the bundle $\Si_{\hat\rho}$ associated to $L(M)\times_M \Si$ by means of the
spin $3/2$ representation
\BE
{\hat\rho}:\GL(m)\times \Spin(r,s)\times V\arr V: (J,S,\psi)\mapsto S^i_{\>j} \psi^j_\nu
(J^{-1})^\nu_{\>\mu}
\EE

The configuration bundle $\calC_{RS}=\Si_\la\times_M\Si_{\hat\rho}$ is gauge natural and
the RS Lagrangian
\BE
L_{_{RS}}=R^{ab}_{\>\>\mu\nu} e_a^\mu e_b^\mu e\>\ds 
+\bar\psi_\mu\ga_5\ga_a\na_\nu\psi_\rho e^a_\si\ep^{\mu\nu\rho\si}\>\ds
\EE
is covariant w.r.t. generalized gauge tranformations.
The theory can be shown again to be gauge natural (see \cite{SuperTesi}) and again it admits
supersymmetries (see \cite{Cubo}). 
This model is very popular in physical applications since the spin frame
arises in it as a gauge field of supersymmetries, so that gravity can be regarded in this case as
a by-product of supersymmetry.

The Rarita-Schwinger model is currently under investigation.
In particular we aim to show that supersymmetries can be regarded as
automorphisms of a suitably defined principal superbundle.
In this case one has a super-principal bundle to which the configuration bundle is associated as
a {\it super-gauge natural bundle}.
The Rarita-Schwinger model will be thence regarded as the prototype of {\it graded gauge natural
theories} which suitably extend the notion of gauge natural theory to encompass supersymmetries
(see \cite{SuperTesi}).

\end{document}